\documentclass[11pt,a4paper]{article}
\usepackage[utf8]{inputenc}
\usepackage{amsmath,amssymb,amsfonts}
\usepackage{colortbl}
\usepackage{geometry}
\usepackage{cancel}

\begin{document}

\title{\textbf{A novel extension of the canonical 2+1-dimensional Dym equation. Moving boundary problems solvable via  Painlev\'e II Reduction}}
\author{
    Colin Rogers$^1$ Sandra Carillo$^{2, 3}$ \footnote{\bf corresponding author,   email: sandra.carillo@uniroma1.it} \\
    \small $^1$University of New South Wales, Sydney, New South Wales, Australia \\
    \small $^2$ Dipartimento Scienze di Base e Applicate per l'Ingegneria \\
    \small Universit\`a di Roma  ``La Sapienza'', 16, Via A. Scarpa, 00161 Rome, Italy \\
    \small $^3$ Gr. Roma1, IV - Mathematical Methods in NonLinear Physics  (MMNLP)\\
    \small National Institute for Nuclear Physics (I.N.F.N.), Rome, Italy
}
\date{\today}

\maketitle

\begin{abstract}
An extension with temporal modulation is presented of the canonical
$2+1$-dimensional solitonic Dym equation as originally introduced by
Konopelchenko and Dubrovsky. Application is made of admitted Painlev\'e II
symmetry reduction to derive exact solution to a class of associated
$2+1$-dimensional nonlinear moving boundary problems of generalized Stefan-type.
\end{abstract}

\section{Introduction}

Moving boundary problems and their reciprocal associates were originally
analysed for the solitonic $1+1$-dimensional Dym equation in \cite{1}.
The physical motivation resided in the occurrence of the canonical Dym
equation in Hele-Shaw problems \cite{2, 3}. In geometric terms, extended Dym
equations have been derived in \cite{4} in connection with binormal spatial
motions of curves with constant curvature. Solitonic surfaces are thereby
generated while the torsion $\tau$ of the curves is governed by the
nonlinear evolution equation
\begin{equation}
\frac{\partial \tau}{\partial b}=
\frac{\partial}{\partial s}
\left[
\frac{1}{\kappa}
\left(\frac{1}{\tau^{1/2}}\right)_{ss}-\tau^{3/2}
+\kappa\left(\frac{1}{\tau^{1/2}}\right)
\right]
\tag{1.1}
\end{equation}
in appropriate arc-length $s$ and binormal $b$ coordinates. Incorporated
as a specialisation of (1.1) on suitable scaling and limiting processes
is an extended Dym equation as originally set down in a hydrodynamic
context by Camassa and Holm \cite{5}, namely
\begin{equation}
u_t+2\partial_x(1-\partial_x^2)
\left(\frac{1}{u^{1/2}}\right)
=0 .
\tag{1.2}
\end{equation}
as derived in a seminal study of novel peakon phenomena in an analysis
of shallow water wave propagation. In \cite{6}, classes of moving boundary
problems for (1.2) and reciprocal associates were subsequently shown to
admit exact solution in terms of Yablonski--Vorob'ev polynomials via
Painlev\'e II symmetry reduction.

That the general torsion equation (1.1) is solitonic was established in
\cite{4} via its connection by a reciprocal--B\"acklund transformation \cite{7}
to the m$^2$KdV equation with origin in work of Fokas \cite{8}. In
addition, the torsion evolution equation (1.1) was shown in \cite{4} to be
embedded in a wide class of nonlinear evolution equations involving
reciprocal B\"acklund transformations of a type delineated in \cite{9}.

A $2+1$-dimensional extension of the canonical $1+1$-dimensional Dym
equation was originally set down by Konopelchenko and Dubrovsky \cite{10}.
In \cite{11}, this novel extension was subsequently shown to be linked by
a $2+1$-dimensional reciprocal transformation to the singularity manifold
equation associated with the application of a Painlev\'e test to the
Kadomtsev--Petviashvili equation descriptive of certain long wave
propagation in hydrodynamics. 

In recent work by Konopelchenko et al. \cite{12} variants of a
$2+1$-dimensional extension of the Camassa--Holm equation have been
investigated with Lax pair and associated $\bar{\partial}$-dressing
schemes detailed. Integrable modulated versions of the $2+1$-dimensional
extended Dym equation were generated via application of a class of
involutory transformations with genesis in classical Ermakov theory \cite{13}.

Here, a novel class of moving boundary problems for an extended
$2+1$-dimensional Dym equation with temporal modulation is shown to admit
exact solution via Painlev\'e II symmetry reduction. Generalisation to moving boundary problems for 2+1 dimensional linked solitonic hierarchies such as set down in \cite{18, 11b, 11c} remains to be investigated.

\section{A Painlev\'e II Symmetry Reduction}

The canonical $2+1$-dimensional Dym equation
\begin{equation}
r_t+r^3r_{xxx}+\frac{3}{r}\left(r^2\partial_x^{-1}
\left(\frac{r_y}{r^2}\right)\right)_y=0
\tag{2.1}
\end{equation}
as originally introduced by Konopelchenko and Dubrovsky in \cite{10}, with
$r=p ^{-1/2}$ becomes
\begin{equation}
p _t+2(p ^{-1/2})_{xxx}
-6p ^2\left(\frac{1}{p }\partial_x^{-1}\partial_y
(p ^{-1/2})\right)_y=0 .
\tag{2.2}
\end{equation}

Here, a novel extended version with temporal modulation of the latter,
namely
\begin{equation}
p _t+2(p ^{-1/2})_{xxx}
-6p ^2\left(\frac{1}{p }\partial_x^{-1}\partial_y
(p^{-1/2})\right)_y(t+a)^k
+2\delta(p^{1/2})_x(t+a)^k=0
\tag{2.3}
\end{equation}
with $k=-2/3$ will be shown to admit a Painlev\'e II symmetry reduction via the ansatz
\begin{equation}
p^{-1/2}=(t+a)^mP(\xi),\qquad
\xi=\frac{x+\alpha^*y}{(t+a)^n},\qquad \alpha^*\in\mathbb{R}.
\tag{2.4}
\end{equation}
Insertion of the latter into (2.3) yields
\begin{align}
&-2m(t+a)^{-2m-1}P^{-2}
-2(t+a)^{-2m-1}(-n\xi P^{-3}P')
+2(t+a)^{m-3n}P''' \notag\\
&\quad+\left[-6(t+a)^{-4m}P^{-4}
\left\{(t+a)^{2m}P^2\partial_x^{-1}\partial_y
\bigl[(t+a)^mP\bigr]\right\}_y\right](t+a)^k \notag\\
&\quad+2\delta(t+a)^{k-m-n}(-P'/P^2)=0 .
\tag{2.5}
\end{align}
wherein
\begin{equation}
\partial_x^{-1}\partial_y\left[(t+a)^mP\right]
=\alpha^*\partial_x^{-1}\left[(t+a)^{m-n}P'\right]
=\alpha^*(t+a)^mP .
\tag{2.6}
\end{equation}
so that
\begin{align}
&(t+a)^{-4m}P^{-4}
\left\{(t+a)^{2m}P^2\partial_x^{-1}\partial_y
[(t+a)^mP]\right\}_y\notag\\
&\qquad=\alpha^{*2}(t+a)^{-2m}P^{-4}
(t+a)^m(t+a)^{-n}(3P^2P')\notag\\
&\qquad=3\alpha^{*2}(t+a)^{-(m+n)}
\frac{P'}{P^2}.
\tag{2.7}
\end{align}

Thus, with
\begin{equation}
m-n=-1/3,\qquad k+m-n+1=0
\tag{2.8,  2.9}
\end{equation}
so that $k=-2/3$, reduction of (2.5) is obtained to
\begin{equation}
P'''=mP^{-2}-n\xi P^{-3}P'
+\left(\delta-9\alpha^{*2}\right)P^{-2}P'.
\tag{2.10}
\end{equation}
Integration of the latter yields
\begin{equation}
PP''-\frac12P'^2-n\xi P^{-1}
+\delta^*
=I+(m-n)\int P^{-1}(\xi)\,d\xi,\qquad I\in\mathbb{R},
\tag{2.11}
\end{equation}
with
\begin{equation}
\delta^*=\delta-9\alpha^{*2}.
\end{equation}

On setting
\begin{equation}
w=aP_\xi,\qquad
s=I+(m-n)\int P^{-1}(\xi)\,d\xi
\tag{2.12,2.13}
\end{equation}
there results
\begin{equation}
w_s=aP''P/(m-n), ~~ w_{ss}=a(m-n)^2\left[
m-\cancel{n\xi P^{-1}P'} + \cancel{\delta^* P'}
+\frac12P^{1/3}+\cancel{n\xi P^{-1}P'}- \cancel{\delta^* P' }+P' s\right],
\tag{2.14}
\end{equation}
whence , on appropriate scalings, the canonical Painlev\'e II equation
\begin{equation}
w_{zz}=2w^3+zw+\alpha,\qquad \alpha\in\mathbb{R}
\tag{2.15}
\end{equation}
is obtained.

\section{A Class of Moving Boundary Problems}

Here, a novel class of moving boundary problems for the extended
$2+1$-dimensional Dym equation (2.3) with temporal modulation
corresponding to $k=-2/3$ is considered so that Painlev\'e II symmetry
reduction is admitted. Thereby, exact solution of the determining
boundary conditions may be obtained.

The boundary conditions adopted are
\begin{align}
&2(p^{-1/2})_{xx}
+\partial_x^{-1}\left[-6p^2\left(\frac1p
\partial_x^{-1}\partial_y(p^{-1/2})\right)_y\right]
(t+a)^{-2/3}
+2\delta p^{1/2}(t+a)^{-2/3}
=L_mS^j\dot S,\quad t>0,
\tag{3.1}\\
&p=P_mS^j,\qquad
\tag{3.2}
\end{align}
on $x+\alpha^*y=S(t),\quad t>0$, with $S(t)=\gamma(t+a)^n$,

\begin{align*}
&2 (p^{-1/2})_{xx} + \partial_x^{-1} \left[ -6 p^2 \left( \frac{1}{p} \partial_x^{-1} \partial_y p^{-1/2} \right)_y \right] (t+a)^{-2/3} + \hfill \\ & \hskip2cm 2 \delta p^{-1/2} (t+a)^{-2/3} \Big|_{x+\alpha^* y = 0} = H_0 (t+a)^k, \quad t > 0 
\tag{3.3}
\end{align*}
The constant parameters $L_m$, $P_m$ and $H_0$ together with the indices $i,j,k$ are determined by admittance of the symmetry ansatz (2.4).

\goodbreak

\begin{center}
{\bf Boundary Conditions Symmetry Reduction}
\end{center}
\begin{description}
    \item[I]  

\begin{align*}
&2(p^{-1/2})_{xx}
+\partial_x^{-1}\left[-6p^2\left(\frac1p
\partial_x^{-1}\partial_y(p^{-1/2})\right)_y\right]
(t+a)^{-2/3}
+2\delta p^{-1/2}(t+a)^{-2/3}
=L_mS^i\dot S,\quad t>0,
\end{align*}
on $x+\alpha^*y=S(t)$, $t>0$.

On insertion of the symmetry representation (2.4) there results

\begin{equation}
2(t+a)^{m-2n}P''(\gamma)
-18\alpha^{*2}(t+a)^{m-n}\left(\frac1{P(\gamma)}\right)
+2\delta(t+a)^{-m-2/3}\frac1{P(\gamma)}
=L_mS^i\dot S,
\tag{3.4}    
\end{equation} 
whence, with $m-n=-1/3$,
\begin{equation}
2\left[P''(\gamma)+\delta^*\left(\frac1{P(\gamma)}\right)\right]
=L_mS^i\dot S(t+a)^{2n-m}.
\tag{3.5}
\end{equation}
so that
\begin{equation}
L_m=2\left[P''(\gamma)+\delta^*(1/P(\gamma))\right]
\frac1{n\gamma^{i+1}}
\tag{3.6}
\end{equation}
together with
\begin{equation*}
    i= \frac{(m-3n+1)}{n}.
\end{equation*} 

\item[II] 
\begin{equation}
p=P_mS^j\quad\hbox{on}\quad x+\alpha^*y=S(t),\quad t>0.
\tag{3.7}
\end{equation}
This condition requires
\begin{equation}
P_m=\gamma^{-j}P^{-2}(\gamma)
\tag{3.8}
\end{equation}s
together with
\begin{equation}
j={2m}/{n}.
\tag{3.9}
\end{equation}

\item[III] 

\item\begin{align*}
&2(p^{-1/2})_{xx}
+\partial_x^{-1}\left[-6 p^2\left(\frac 1 p 
\partial_x^{-1}\partial_y(p^{-1/2})\right)_y\right]
(t+a)^{-2/3} \\
&\qquad +2\delta p^{1/2}(t+a)^{-2/3}\vert_{x+\alpha^* y=0}
=H_0 (t+a)^k,\quad t>0,
\end{align*}

This yields
\begin{equation}
H_0=2\left[P''(0)+\delta^*P^{-1}(0)\right],
\tag{3.10}
\end{equation}
together with
\begin{equation}
k=m-2n.
\tag{3.11}
\end{equation}
\end{description}

The present $2+1$-dimensional moving boundary problems as in the
case of the $1+1$-dimensional Dym and $1+1$-dimensional Camassa--Holm
type extension admit Yablonski--Vorob'ev polynomial solutions as
admitted by the Painlev\'e II symmetry reduction. Nonlinear moving
boundary problems of Stefan-type for both the canonical Korteweg-deVries  
and mKdV solitonic equations have previously been shown in \cite{14, 15}, 
respectively, to admit exact solution
via Painlev\'e II symmetry reduction involving the classical Airy equation.
This corresponds to the specialisation of the Painlev\'e II equation (2.15)
with $\alpha=-1/2$ and
\begin{equation}
w=-\frac{\phi'(z)}{\phi(z)}
\tag{3.12}
\end{equation}
with Airy equation
\begin{equation}
\phi''+\frac{z}{2}\phi=0.
\tag{3.13}
\end{equation}

It is remarked that, application of a B\"acklund transformation originally
due to Lukashevich \cite{16} to this Airy class generates exact solutions of
Painlev\'e II corresponding to a sequence of Airy parameters. This result which may be applied in the present
moving boundary context has previously been applied in a physical
context \cite{17} to solve novel classes of stationary boundary value
problems for the Nernst-Planck  electrolytic system.

\subsection*{Acknowledgements}

S.C.\ acknowledges:
\begin{itemize}
\item Gr. Roma1, IV - Mathematical Methods in NonLinear Physics (MMNLP), National Institute for Nuclear Physics (I.N.F.N.), Rome, Italy;
\item Gruppo Nazionale di Fisica Matematica (G.N.F.M- I.N.D.A.M.), Italy;
\item Regione Lazio,   project CTE, Italy
\end{itemize}
 for support of current activity.

\end{document}